\documentclass[a4paper]{llncs}
\usepackage{graphicx}
\usepackage{amsmath}
\usepackage{url}
\usepackage{hyperref}
\begin{document}
\title{Throttling Malware Families in 2D}
\author{Mohamed Nassar \and Haidar Safa
}
\institute{Department of Computer Science \\
American University of Beirut \\
Beirut, Lebanon \\
\email{[mn115|hs33]@aub.edu.lb}\\
}
\maketitle              % typeset the header of the contribution
\begin{abstract}
Malicious software are categorized into families based on their static and dynamic characteristics, infection methods, and nature of threat. Visual exploration of malware instances and families in a low dimensional space helps in giving a first overview about dependencies and relationships among these instances, detecting their groups and isolating outliers. Furthermore, visual exploration of different sets of features is useful in assessing the quality of these sets to carry a valid abstract representation, which can be later used in classification and clustering algorithms to achieve a high accuracy.  In this paper, we investigate one of the best dimensionality reduction techniques known as t-SNE to reduce the malware representation from a high dimensional space consisting of thousands of features to a low dimensional space. We experiment with different feature sets and depict malware clusters in 2-D.  Surprisingly, t-SNE does not only provide nice 2-D drawings, but also dramatically increases the generalization power of SVM classifiers. Moreover, obtained results showed that cross-validation accuracy is much better using the 2-D embedded representation of samples than using the original high-dimensional representation.

\end{abstract}
\section{Introduction}
Security breaches are executed through malwares and are a major threat to the Internet today. There are several forms of malware ranging from viruses and spam bots to trojan horses and rootkits \cite{stallings2016cryptography}. Recently, the Petya ransomware \cite{akkas2017malware} crashed shipping companies, ports, and law agencies. This malware targets the master boot record of a machine and prohibits the operating system from normal execution. It then spreads and encrypts all the system files. A message appears on the screen stating the amount of ransom to decrypt the files. The payment is through crypto-currencies. The prominent expansion of malwares is due to their metamorphic and polymorphic techniques that give the ability to change their code as they propagate. In addition, malwares adopt new ways to detect the environments where they are running, hence hindering their detection and making dynamic analysis difficult if not impossible. 

Visual analytics provides approaches to obtain an understanding from complex data. It aims at developing methods that allow analysts to examine the processes underlying the data \cite{ellis2010mastering}. Visual exploration of malware families is a pre-processing step of a more in-depth malware family analysis, as it allows for the development of intuitions and hypotheses about the discriminative power of a set of contextual or behavioral features.  However, visualizing malware families in low dimensional space (2-D, 3-D) is a topic that received little attention in the literature. Malware data are fundamentally different than text and images\footnote{\url{http://www.jsylvest.com/blog/2017/12/malconv/}} which motivates investigating ways to adapt existing approaches or inventing new ones. For instance a byte in malware has different meanings in different contexts, in contrast to a byte representing pixel intensity in an image. 

In this paper, we experiment with the best low-dimensional embedding technique known as t-SNE (Student-t distribution -- Stochastic Neighborhood Embedding) for depicting malware clusters and features. We propose a pipeline for feature extraction and selection, followed by visualization. We compare the raw classification accuracy at the high-dimensional and low-dimensional spaces for n-grams features. Finally We propose a new first-insight classifier based on t-SNE and SVM. Note that our goal is not to propose a very high accuracy classifier or to compete with extensive feature selection approaches. Instead, we aim at exploring the visualization space of malware families and to which extent such a pre-processing procedure might be useful for analyzing a typical malware dataset. The remaining of the paper is organized as follows. Section 2 surveys relevant related work. Section 3 presents our proposed methodology. In Section 4, we discuss implementation and results. We finally conclude in Section 5 and present future research directions.

\section{Background and Related Work}
Classifying and Clustering malware families were addressed in many recent work in the literature. In \cite{li2010challenges}, the current automated approaches for malware clustering were summarized. The paper considered the high accuracy obtained by six commercial anti-viruses as biased since unbalanced datasets were used where most malware instances are easy to classify. A plagiarism detector algorithm was applied on the same dataset and yielded the same accuracy results compared to those of the anti-viruses, though the plagiarism detector does not have any expert knowledge about malwares. In \cite{bayer2009scalable} a scalable, behavior-based malware clustering approach was proposed. This approach aims to isolate outliers that exhibit a novel behavior to be further analyzed. It used a recent technique called taint tracking to build behavioral profiles and locality sensitive hashing for clustering these profiles. Malware instance visualization was proposed in \cite{nataraj2011malware}. This approach suggested transforming the binary into a vector of 8-bit integers, which can be reshaped into a matrix and therefore viewed as a gray-scale image. This technique proves useful in increasing the accuracy of malware classifiers. The work presented in this paper is different such that we focus on visualizing families of malwares as scatter plots using t-SNE \cite{maaten2008visualizing}, an embedding technique that allows visualizing high-dimensional data by giving each datapoint a location in a two or three-dimensional map. The technique is a variation of Stochastic Neighbor Embedding (SNE). 

SNE starts by converting the high-dimensional Euclidean distances between datapoints into conditional probabilities that represent similarities. The similarity of datapoint $x_j$ to datapoint $x_i$ is the conditional probability, $p_{j|i}$ , that $x_i$ would pick $x_j$ as its neighbor if neighbors were picked in proportion to their probability density under a Gaussian centered at $x_i$. For the low-dimensional counterparts $y_i$  and $y_j$ of $x_i$ and $x_j$, it is possible to compute a similar conditional probability $q_{j|i}$. SNE aims to find a low-dimensional data representation that minimizes the mismatch between $p_{j|i}$ and $q_{j|i}$. To do so, SNE minimizes the sum of Kullback-Leibler divergences over all datapoints using a gradient descent method. t-SNE differs from the old SNE in two ways: (1) it uses a symmetrized version of the SNE cost function with simpler gradients and (2) it uses a heavy-tailed Student-t distribution rather than a Gaussian distribution to compute the similarity between two points in the low-dimensional space. This allows t-SNE to alleviate both the so-called crowding problem and the optimization problems of SNE \cite{maaten2008visualizing}.

The dataset we use is the training set from Microsoft Malware Classification Challenge \cite{ronen2018microsoft} released in 2015, which since has been studied by researchers in more than 50 publications targeting feature engineering, deep learning, clustering and classification approaches. For instance, \cite{gibert2016convolutional} proposes using convolutional neural networks for feature extraction and classification based on the binary and reconstructed assembly files. Our work takes another direction and focuses on the visualization question.

\section{Proposed Methodology}
In this section we describe our proposed methodology which copes with relatively large data sets. We have used the training dataset from Microsoft malware classification challenge (Big 2015) having 10868 labeled instances \cite{ronen2018microsoft}. The standard version of t-SNE having quadratic complexity in terms of the number of instances $O(N^2)$ might be applied. For larger data-sets, other versions of t-SNE are proposed such as random-walk based sampling of landmark points or using specialized data structures leading to  $O(N \log N)$ complexity \cite{van2014accelerating}. We particularly used the version of scikit-learn with the Barnes-Hut approximation running in $O(N \log N)$ time.

\begin{figure}[htb]
\centering
\includegraphics[width=\textwidth]{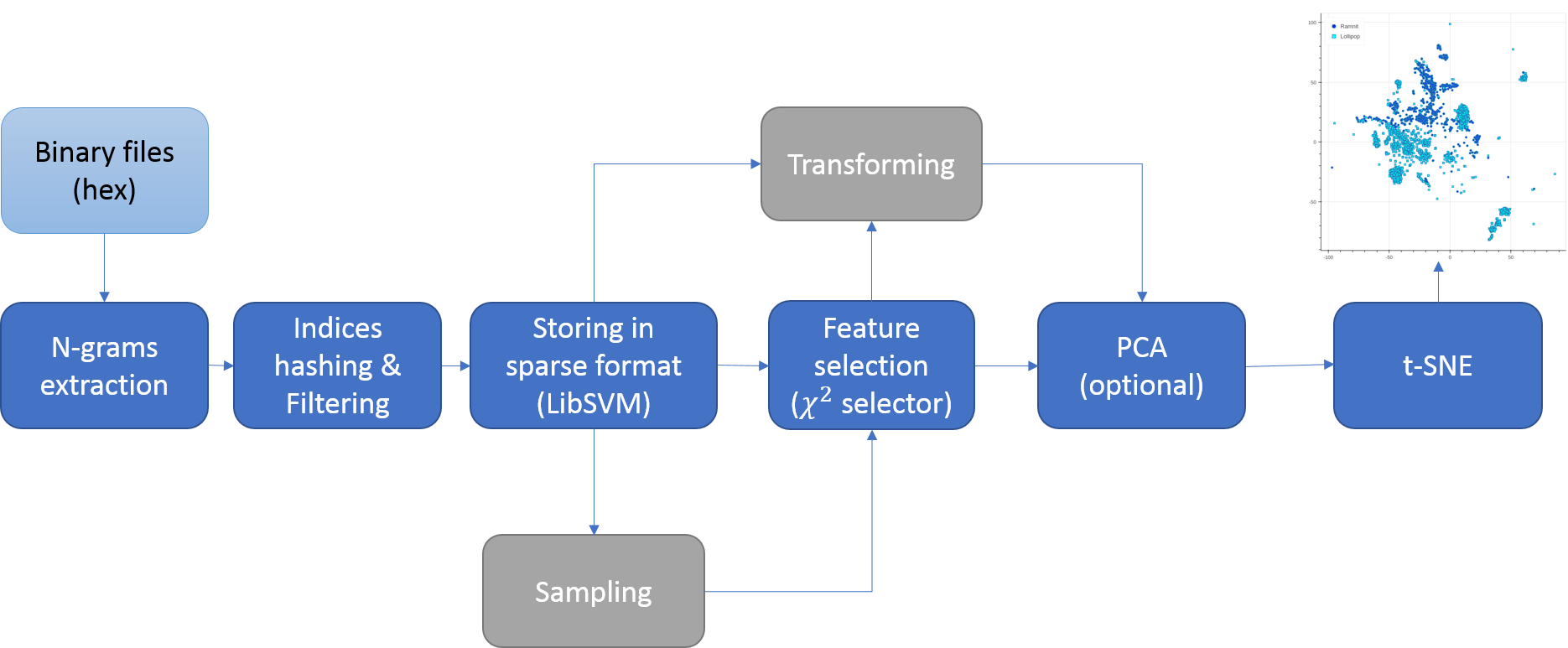}
\caption{Pipeline for malware family visualization.} \label{fig1}
\end{figure}

Fig.~\ref{fig1} shows the pipeline of our proposed methodology. Starting from a labeled corpus of files containing the malwares payload in hexadecimal format, we extract n-byte grams (3, 4 and 5). Note that the number of features grow exponentially such that for 3-byte grams we have $2^{8*3}$ possible words, for 4-byte grams it is $2^{8*4}$ and so on. Most current machine learning libraries cannot handle this number of features even in sparse format. For example, Scikit-learn accepts feature indices less than a positive 4-bytes signed integer ($2^{31}-1$). Therefore, in the proposed methodology, we hash the feature indices to 22-bits integers. We also proceed by early removal of rare words that appear less than k times (we assumed k = 3). These words probably represent addresses in memory or literals and have little differentiation power. This technique is very efficient in reducing the storage amount of the features set. We store the resulting features along with the instance label in sparse format (LibSVM/SVMLight format) as one line per instance in the output text file. Each feature is represented by an index (the hash of the n-bytes gram) and a value which is the number of occurrences of this n-bytes gram in the malware instance.  This stage is implemented by using the Sally tool \cite{rieck2012sally} which is an efficient feature extraction tool that generates n-grams besides other features such as TF-IDF \cite{chowdhury2010introduction}.

We proceed with a first feature selection stage using the $\chi^2$ statistical test-based selector where our target is to reduce the number of features to 1,000. This limit reduces the complexity of computing the pair-wise distances in t-SNE. However, we are starting with a much larger number ($2^{22} = 4,194,304$ feature-space). The time complexity of the $\chi^2$ selector is $O(n_\text{classes} * n_\text{features}$). The features that are the most likely to be independent of class and therefore irrelevant for classification are removed. 

We also can reduce the space complexity of this stage (mainly because of memory limitations) by sampling the instances in equal proportions to their family sizes. If sampling is used than we must use the generated $\chi^2$ selector model to transform the complete dataset and keep the top $1,000$ features of each instance. For the dataset in question we did not use sampling. 

Optionally, we apply a PCA (Principal Component Analysis) transformation to reduce further the number of features to the range of 30-50 features. This speeds up the computation of pairwise distances between the data-points in the next stage and suppresses some noise without severely distorting the inter-point distances \cite{maaten2008visualizing}. t-SNE then embeds these features in 2 dimensions. The malware instances are depicted as scatter plots. t-SNE outperforms other data embedding techniques such as PCA, Sammon mapping, Isomap and LLE. 
\section{Implementation, Preliminary Results, and Interpretation
}
In this section, we describe the implementation environment and setup, the dataset used, the hyper parameters then we discuss the obtained results. 
\subsection{Setup, implementation, and tools}
All our experiments were performed using a commercial off-the-shelf laptop with a 64-bit Ubuntu 16.04 LTS operating system, an Intel core i5-5200U CPU (4 cores, 2.20GHz) 8 GB RAM and 1 TB Hard disk. Our implementation was based on Python v3.6, numpy v1.13, scipy v0.19 and the Scikit-learn v0.19 library. The plots are generated using the BokehJS v0.12 library. We have used Sally \cite{rieck2012sally} for feature extraction. Our code is available at \url{https://github.com/mnassar/malware-viz} under form of Python Jupyter notebooks for further exploration and result reproducibility.
\subsection{Dataset}
The dataset is the training set from Microsoft Malware Classification Challenge \cite{ronen2018microsoft}, which includes 10868 labeled samples. For each sample, the raw data and meta data are provided. The raw data contains the hexadecimal representation of the file's binary content, without the Portable Executable (PE) header to ensure sterility. The metadata manifest is a log containing various metadata information extracted from the binary, such as function calls, strings, etc. This was generated using the IDA disassembler tool. In our implementation, we focused solely on the raw data. However, we consider augmenting our visualization with the meta data for future work. The raw training data volume is about 41 GBytes. However, our feature selection method reduces this to about 6.4 GBytes for the sum of three feature sets (3, 4 and 5-grams). The dataset contains malwares belonging to the following 9 families: Ramnit, Lollipop, Kelihos Ver. 3, Vundo, Simda, Tracur, Kelihos Ver. 1, Obfuscator.ACY and Gatak. One challenge of this data set is the unbalanced sizes of different families. The distribution of instances for the training dataset is shown in Table~\ref{tab1}. It is shown that classes 4, 5, 6 and 7 are underrepresented as compared to the other families. 

\begin{table}
\centering
\caption{Distribution of samples among the families.}\label{tab1}
\begin{tabular}{|c|l|l|c|c|}
\hline
Class & Family & Type & Nb. Of Instances & Percentage (\%) \\ 
\hline
1 & Ramnit & Worm & 1541 & 14.20 \\
2 & Lollipop & Adware & 2478 & 22.80 \\
3 & Kelihos\_ver 3 & Backdoor & 2942 & 27.07 \\
4 &  Vundo & Trojan & 475 & 4.37 \\
5 & Simda & Backdoor & 42 & 0.39 \\
6 & Tracur & Trojan Downloader & 751 & 6.91 \\
7 & Kelihos\_ver 1 & Backdoor & 398 & 3.66 \\
8 & Obfuscator.ACY & Obfuscated malware & 1228 & 11.30 \\
9 & Gatak & Backdoor & 1013 & 9.32 \\
\hline
\end{tabular}
\end{table}

\subsection{t-SNE parameters}
t-SNE implementation in Sklearn has many tunable parameters. We show the most important ones with a short description in Table~\ref{tab2}. Note that the t-SNE method is known to be little sensitive to these parameters. Nevertheless, a good tuning enhances the quality of the obtained clusters sometimes. 

\begin{table}
\centering
\caption{Tunable Parameters in t-SNE.}\label{tab2}
\begin{tabular}{|c|p{6cm}|p{3.5cm}|}
\hline
\textbf{Parameter} & \textbf{Description } & \textbf{Typical/default value} \\
\hline
\textit{Random\_state} &
This is the seed of random initialization in the embedded space. It is an important parameter to accomplish comparisons under same initial conditions (used with init = 'random') &
No typical value. We have chosen 42 as our default. \\
\hline
\textit{N\_iter} &
Maximum number of iterations for the optimization. &
Default: 1000 \\
\hline
\textit{Perplexity} &
The perplexity can be interpreted as a smooth measure of the effective number of neighbors. &
5 -- 50
Our default is 40 \\
\hline
\textit{Early exaggeration} &
Larger values of this parameter tend to start with distant clusters in the embedded space &
Default: 12 \\
\hline
\textit{metric} &
The distance between instances where each instance is a feature array.  &
Euclidean \\
\hline
\textit{Learning rate (lr) }&
It must not be too low or too high. If the cost function gets stuck in a bad local minimum increasing the learning rate may help. &
10 -- 1000 
We have chosen 200 as our default. \\ 
\hline
\end{tabular}
\end{table}

\subsection{Sample of results}
\begin{figure}
\centering
\includegraphics[width=0.7\textwidth]{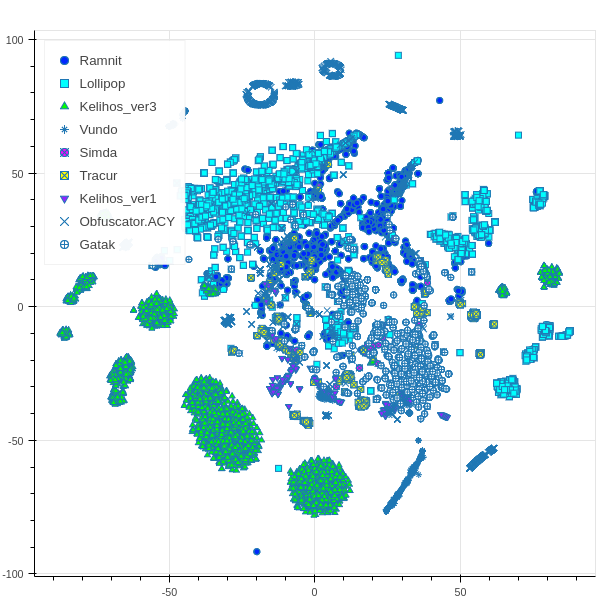}
\caption{4-grams, no-PCA, 9 classes, perplexity=40, lr=200.} \label{fig2}
\end{figure}

Due to lack of space we do not show all obtained visual plots. However, we depict and interpret the most important ones. Fig.~\ref{fig2} shows the scatter plot of the 9 classes for 4-byte grams (no PCA). In this figure, it is directly noticed that each family is composed of one or multiple clusters. No one clear cluster per family exists as in the case of the MNIST dataset \cite{lecun1998gradient}. This is expected due to the polymorphic nature of malware data which has much more noise than what can be carried by handwritten digits or letters. Common evasion and obfuscation techniques may also explain why some families have intersection regions in between their clusters. Still we can identify big clusters for each of the big families (1, 2, 3, 8, 9) and even some smaller clusters for the families 4 (Vundo) and 7 (Kelihos\_ver 1). It is also important to notice the existence of some outliers.

\begin{figure}
\centering
\includegraphics[width=0.7\textwidth]{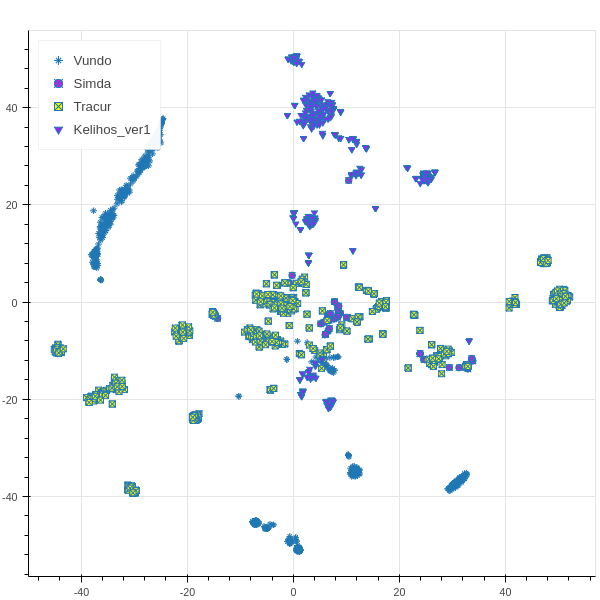}
\caption{4-grams, no-PCA, classes 4,5,6,7, perplexity=40, lr=200.} \label{fig3}
\end{figure}

\begin{figure}
\centering
\includegraphics[width=0.48\textwidth]{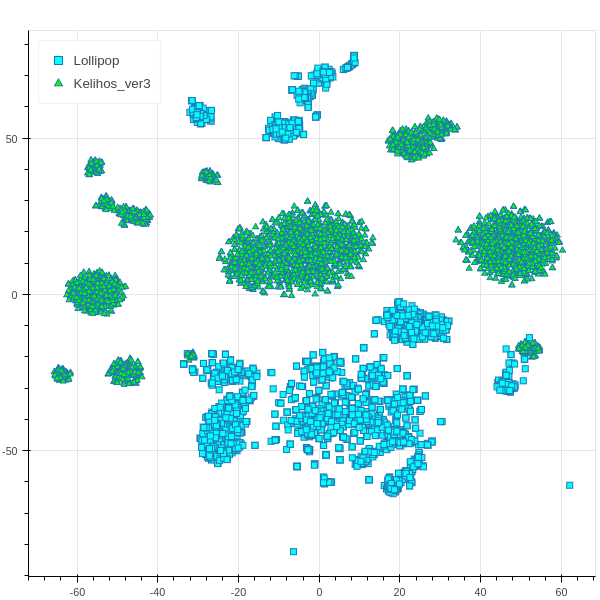} \hspace{0.2cm}   
\includegraphics[width=0.48\textwidth]{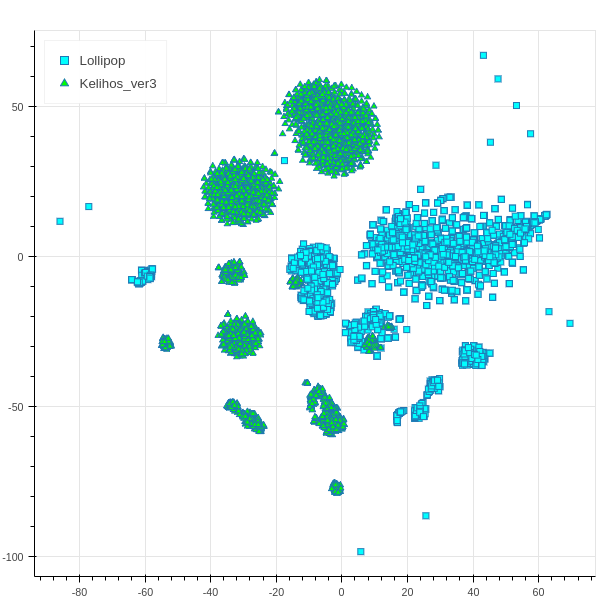}  \\
(a) \hspace{6cm} (b) \\
\vspace{0.1cm}
\includegraphics[width=0.48\textwidth]{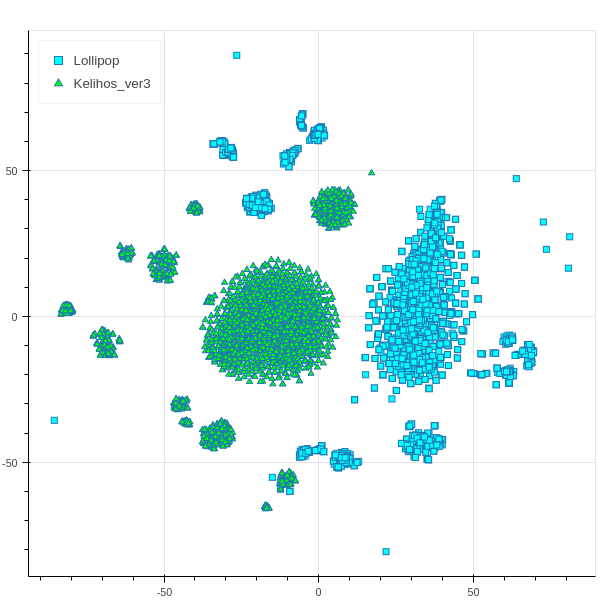} \\
(c) 
\caption{(a) Classes 2 and 3   (3-grams, perplexity=40, lr=200); (b) Classes 2 and 3 (4-grams, perplexity=40, lr=200); (c) Classes 2 and 3 (5-grams, perplexity=40, lr=200)
} \label{fig4}
\end{figure}

To give a chance to the underrepresented families, we started over with only their instances in the pipeline. Obtained results are shown in Fig.~\ref{fig3}. These results are much better, and we can observe clear groups for each of these small families. Next, we take only the largest two classes (2 and 3) to compare plots between different sets of features (3-bytes grams, 4-bytes grams and 5-bytes grams). Obtained results are illustrated in Fig.~\ref{fig4}. They show that to some extent, 5-grams (Fig.~\ref{fig4}(c)) have less overlap between clusters than 3-grams (Fig.~\ref{fig4}(a)) or 4-grams (Fig.~\ref{fig4}(b)). We wanted to correlate this with the classification accuracy on the original dataset (the one after feature selection which is 5420 instances * 1000 features) and the transformed dataset (5420 instances * 2 features). Results in Table~\ref{tab3} shows that training accuracy is almost perfect for all three feature sets. However, training accuracy is often not a good metric since the classifier might overfit the training set. 

We compute the two-fold cross validation accuracy which is much more indicative of the generalization power of the classifier. Astonishingly, the two-fold cross validation accuracy is very poor on the original dataset and dramatically much better on the embedded dataset. This can be explained by the fact that t-SNE groups the datapoints into separate clusters in a low dimensional space, which is at the origin of the design of support vector classifiers with the Radial Basis Function (RBF) kernel. These classifiers shine under this kind of settings. The SVC accuracy (\%) for the three feature sets with default Sklearn parameters (RBF kernel, $C=1.0$, $\gamma=1/n_\text{features}$) are shown in Table~\ref{tab3}. Better SVC accuracy results are usually obtained after a grid search for the best hyperparameters $C$ and $\gamma$.

\begin{table}
\centering
\caption{SVC Accuracy on different feature-sets (classes 2 and 3).}\label{tab3}
\begin{tabular}{|c|c|c|c|c|}
\hline
Feature Set & Training   & Training 
 & Two-fold cross- 
&  Two-fold cross- \\ 
& Accuracy & Accuracy & Validation Accuracy & Validation Accuracy \\
& (1000-d) & (2-d) &   (1000-d) & (2-d) \\
\hline
3-byte grams & 99.98 & 99.98 & 67.91 & 99.57 \\ 
\hline
4-byte grams & 100.00 & 99.96 & 64.06 & 99.88 \\
\hline
5-byte grams & 99.98 & 99.98 & 68.13 & 99.92 \\
\hline
\end{tabular}
\end{table}

\begin{figure}
\centering
\includegraphics[width=0.45\textwidth]{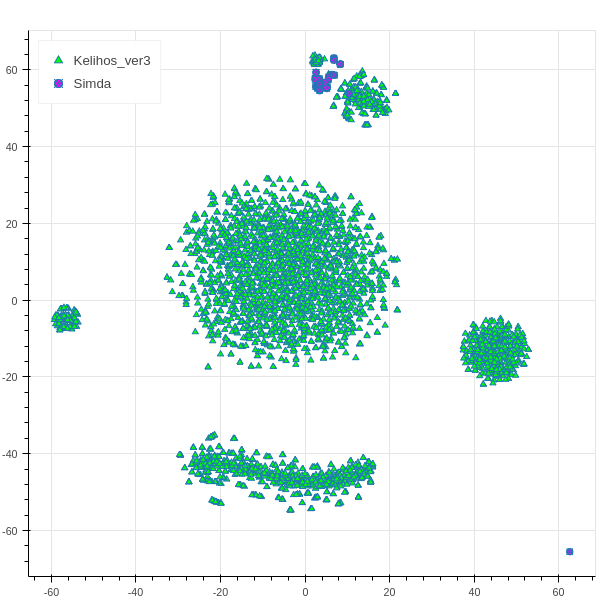}  \hspace{1cm}  
\includegraphics[width=0.45\textwidth]{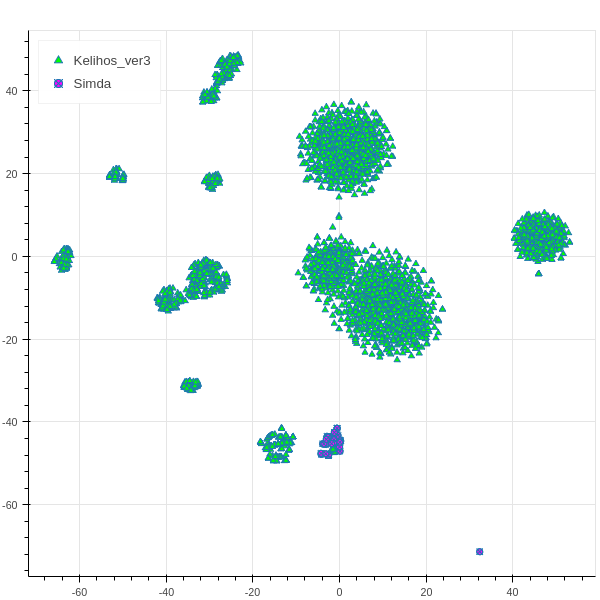} \\
(a) \hspace{5cm} (b) \\
\vspace{0.1cm}
\includegraphics[width=0.45\textwidth]{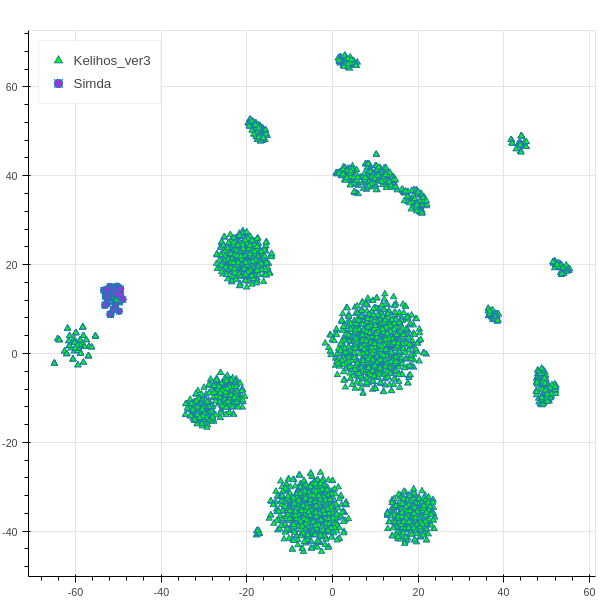} \hspace{1cm}
\includegraphics[width=0.45\textwidth]{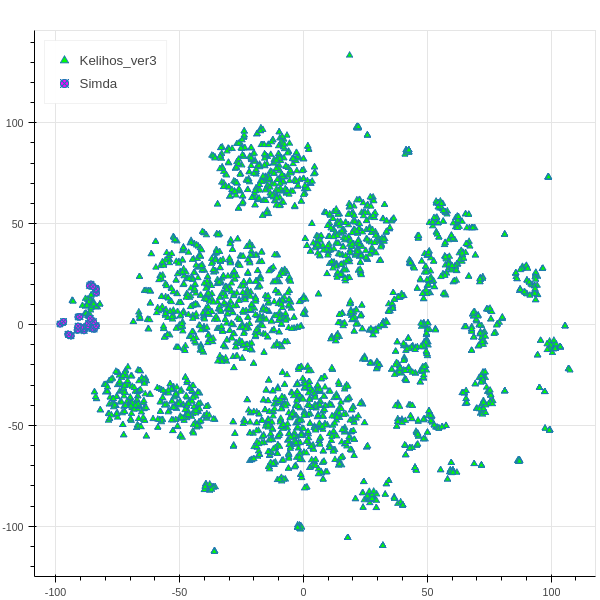} \\ 
(c) \hspace{5cm} (d) \\  
\caption{(a) Classes 3 and 5 (3-grams, perplexity=40, lr=20); (b) Classes 3 and 5 (4-grams, perplexity=40, lr=200); (c) Classes 3 and 5 (5-grams, perplexity=40, lr=200); (d) Classes 3 and 5 (5-grams, perplexity=5, lr=200)
} \label{fig5}
\end{figure}

Next, we examine the performance of t-SNE on unbalanced families. We take the extreme case by choosing the largest family (Kelihos\_ver 3 -- class 3) and the smallest family (Simda -- class 5). Results that are presented in Fig.~\ref{fig5} show that t-SNE can isolate the class 5 in a small cluster. They also show that 4-grams and 5-grams perform better than 3-grams in isolating clusters of the two classes. Fig.~\ref{fig5}(d) shows that choosing a bad perplexity value might degrade the clustering quality. 

t-SNE shows similar performance in enhancing the classification accuracy as shown in Table~\ref{tab4}. Note that since the classes are severely unbalanced, an accuracy of 98.60 would be simply obtained if the classifier considers all the data points a belonging to the majority class 5. The two-fold cross validation accuracies on the original dataset are bad in this sense. However, we notice that the two-fold cross validation accuracy is much better in the embedded space. 3-grams features perform the worst as it can be expected by examining the corresponding scatter plot.

\begin{table}
\centering
\caption{SVC Accuracy on different feature sets (classes 3 and 5)}\label{tab4}
\begin{tabular}{|c|c|c|c|c|}
\hline
Feature Set & Training   & Training  & Two-fold cross- &  Two-fold cross- \\ 
& Accuracy & Accuracy & Validation Accuracy & Validation Accuracy \\
& (1000-d) & (2-d) &   (1000-d) & (2-d) \\
\hline
3-byte grams & 100.00 & 99.93 & 98.52 (bad) & 99.83 \\
\hline
4-byte grams & 100.00 & 99.93 & 98.62 (bad) & 99.90 \\
\hline
5-byte grams & 100.00 & 99.97 & 98.62 (bad) & 99.90 \\
\hline
\end{tabular}
\end{table}

Finally, we want to validate this hypothesis on the complete dataset (9-classes). Results are shown in Table~\ref{tab5}. The training accuracy in 2-d is a bit smaller than in 1000-d but allows much better generalization of the classification model as clearly inferred from the cross-validation accuracy results. 

\begin{table}
\centering
\caption{SVC Accuracy on different feature sets (all classes)}\label{tab5}
\begin{tabular}{|c|c|c|c|c|}
\hline
Feature Set & Training   & Training  & Two-fold cross- &  Two-fold cross- \\ 
& Accuracy & Accuracy & Validation Accuracy & Validation Accuracy \\
& (1000-d) & (2-d) &   (1000-d) & (2-d) \\
\hline
3-byte grams & 
99.66 &
96.84 &
56.76 &
94.26 \\
\hline
4-byte grams &
99.58 &
96.22 &
55.26 &
93.13 \\
\hline
5-byte grams &
98.25 &
95.69 &
60.24 &
92.58 \\
\hline
\end{tabular}
\end{table}
\subsection{Testing Accuracy}
\begin{figure}
\centering
\includegraphics[width=1\textwidth]{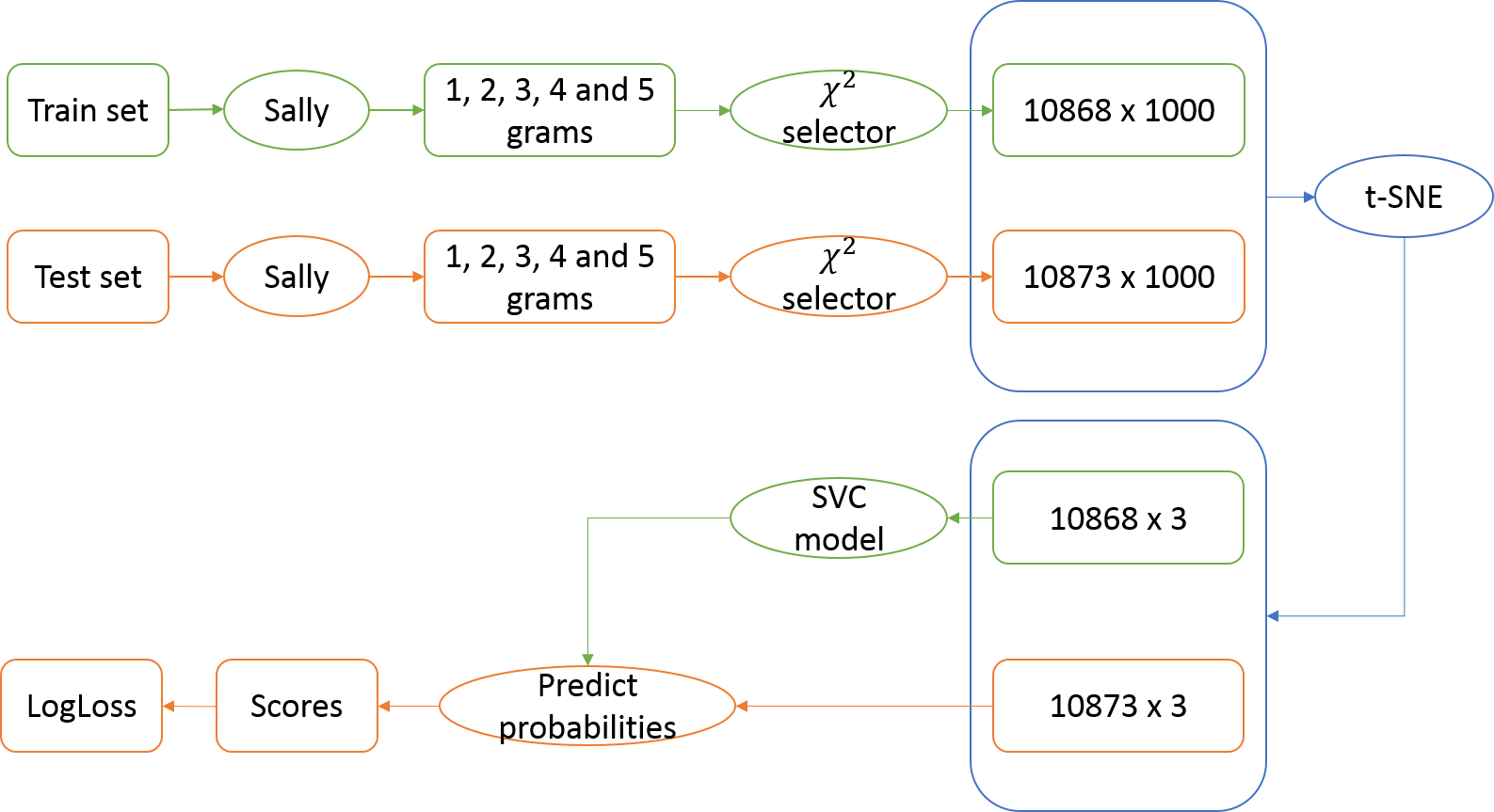} 
\caption{Testing Accuracy Pipeline} \label{fig6}
\end{figure}

In this section we further assess the idea of squeezing the dimensions into a small hyperspace using t-SNE than expanding it back to infinite dimensional space using the RBF kernel with SVM. 
We wanted to estimate the testing accuracy using the raw unlabeled test dataset (10873 instances). Note that t-SNE is a non-parametric mapping, therefore we cannot use the learnt model to map the test datapoints to the embedded space which is formed by the train datapoints. As an alternative we run t-SNE on the full dataset composed of both train and test datapoints (Another approach is to train a multivariate regressor to predict the map location from the input data \cite{van2009learning}). We then fit an SVC model solely based on the train embedded datapoints. This SVC model is used to estimate probabilistic predictions of the membership of each embedded test point to each of the 9 possible classes. The pipeline of this approach is depicted in Fig.~\ref{fig6}. 

Note that the labels of these instances are not available, and the only way of evaluation is to obtain the multi-class logarithmic loss by submitting our predictions in probabilistic form to the dataset hosting platform (Kaggle) online. The equation for the logarithmic loss is: $$\text{logloss}= -\frac{1}{10873}\sum_{i=1}^{10873}\sum_{j=1}^{9}y_{ij}\log p_{ij} $$ where $y_{ij}=1$ if $i$ belongs to class $j$ and $0$ otherwise, $\log$ is the natural logarithm and $p_{ij}$ is the probability that $i$ belongs to class $j$ as given by the classifier.

Our classifier achieves a testing logloss of 0.1719. This is fairly acceptable given the simplicity of the employed feature set (1000 best features among 1,2,3,4 and 5 n-bytes grams) and without recurring to any involved feature engineering. A clueless classifier scores 2.1972. For this experiment we have used an m5.4xlarge AWS EC2 instance (64 GB RAM) and a 500 GB volume.

\section{Conclusion}
In this paper, we have successfully applied feature extraction, selection, embedding and visualization over a recent malware dataset. We have proposed a pipeline that can cope with dataset of similar or larger size. We use the t-SNE algorithm to embed the malware datapoints in 2D and visualize them as scatter plots. A very interesting result is that compressing the data using t-SNE dramatically enhances the cross-validation accuracy of Support Vector Machines classifiers. t-SNE shapes the clusterability of datapoints in the embedded space, which is very appealing to SVM classifiers with the RBF kernel.  

In future work, we aim to experiment with other feature sets, for instance by analyzing the assembly data files. We also want to assess the viability of the SVM--t-SNE classifier over other data sets. Another direction is to work on implementing t-SNE in a 3D WebGL framework and integrate it in Jupyter notebooks. Available GPUs can also be used to bear some of the tedious computations. 

\section*{Acknowledgement}
This work was supported in part by a grant from the
University Research Board of the American University of
Beirut, Lebanon.
\bibliographystyle{splncs04}
\bibliography{malwareViz}
\end{document}